\documentclass[runningheads]{llncs}

\usepackage{graphicx}
\usepackage{amsmath}
\begin{document}
\title{A Deep Discontinuity-Preserving Image Registration Network}
%
\author{Xiang Chen\inst{1} \and 
Nishant Ravikumar\inst{1,2} \and 
Yan Xia\inst{1,2} \and 
Alejandro F Frangi\inst{1,2,3,4}} 
\authorrunning{X. Chen et al.}
%
\institute{Center for Computational Imaging and Simulation Technologies in Biomedicine, School of Computing, University of Leeds, Leeds,UK \and
Biomedical Imaging Department, Leeds Institute for Cardiovascular and Metabolic Medicine, School of Medicine University of Leeds, Leeds, UK\\
\and Department of Cardiovascular Sciences, KU Leuven, Leuven, Belgium\\
\and
Department of Electrical Engineering, KU Leuven, Leuven, Belgium}

%
\maketitle              
\sloppy
\begin{abstract}
Image registration aims to establish spatial correspondence across pairs, or groups of images, and is a cornerstone of medical image computing and computer-assisted-interventions. Currently, most deep learning-based registration methods assume that the desired deformation fields are globally smooth and continuous, which is not always valid for real-world scenarios, especially in medical image registration (e.g. cardiac imaging and abdominal imaging). Such a global constraint can lead to artefacts and increased errors at discontinuous tissue interfaces. To tackle this issue, we propose a weakly-supervised Deep Discontinuity-preserving Image Registration network (DDIR), to obtain better registration performance and realistic deformation fields. We demonstrate that our method achieves significant improvements in registration accuracy and predicts more realistic deformations, in registration experiments on cardiac magnetic resonance (MR) images from UK Biobank Imaging Study (UKBB), than state-of-the-art approaches. 

\keywords{Deep Learning  \and Image Registration \and Cardiac Image Registration \and Discontinuity-preserving Image Registration.}
\end{abstract}

\section{Introduction}
Image registration is a fundamental component of several applications in medical imaging. Recent years have seen a shift from traditional iterative methods to deep learning (DL)-based registration approaches. Although training DL-based approaches is time-consuming, inference is rapid, involving just a single forward pass through the network. Consequently, DL-based approaches offer substantial acceleration for pair-/group-wise image registration relative to traditional approaches, achieving near-real-time performance in certain applications.

Most existing DL-based registration methods constrain deformation fields to be globally smooth and continuous, through various means~\cite{balakrishnan2019voxelmorph,dalca2019unsupervised,krebs2019learning}. However, this assumption is often violated in medical image registration applications, as tissue boundaries are naturally discontinuous. This is especially pronounced in cardiac or abdominal imaging, which involve large deformations of multiple tissue-types, and organ motion/sliding at tissue boundaries. Variability in the physical properties of different tissue-types results in discontinuities at native tissue boundaries~\cite{hua2017multiresolution,hua2016non}. 
Hence, enforcing deformation fields to be globally smooth can generate unrealistic deformations and lead increased errors near these boundaries.

Discontinuity-preserving image registration is an active area of research in the context of traditional registration methods~\cite{wu2008evaluation,schmidt2012estimation,pace2013locally,hua2017multiresolution}. 
For example, Hua et al.~\cite{hua2017multiresolution} proposed a discontinuous registration approach that utilised enriched B-spline basis functions at control points near discontinuous tissue boundaries, achieving significant improvement in registration accuracy, relative to other existing discontinuity-preserving registration methods. 
In contrast, only one study thus far has proposed a discontinuous DL-based image registration framework. Ng et al.\cite{ng2020unsupervised} proposed a custom discontinuity-preserving regulariser on the deformation fields (used with a typical unsupervised registration network), to preserve discontinuities, while ensuring local smoothness within specific regions. They formulated a regularisation term based on the unsigned area of the parallelogram spanned by two displacement vectors associated with moving image voxels. However, without additional boundary information for guidance, such a discontinuity regularisation term alone is insufficient to preserve strong discontinuities in deformation fields.

This paper assumes that the desired deformation fields are locally smooth, but discontinuities may exist between different regions/organs at tissue interfaces. Therefore, we generate distinct smooth deformation fields for different regions of interest and compose them to obtain the final registration field, used to warp the moving image. Such a locally-smooth and globally-discontinuous registration scheme is achieved using a novel Deep Discontinuity-preserving Image Registration network, or DDIR. Our approach obtains significant improvements in registration accuracy relative to the state-of-the-art and preserves discontinuities at tissue/region boundaries in the estimated deformation fields. This yields more realistic deformations than afforded by state-of-the-art methods.

In summary, the contributions of this paper are two-fold: (1) we designed a novel framework, DDIR for discontinuous DL-based image registration. This is the first study to incorporate discontinuity in DL network structure and training strategy, and not only in terms of a custom regularisation term in the loss function.
(2) Our proposed DDIR achieves significant improvement in registration accuracy over state-of-the-art registration methods, and preserves key cardiac morphological indices post-registration, not afforded by the latter.

\section{Method}
Pair-wise image registration aims to establish spatial correspondence between the moving image $\textbf{I}_M$ and fixed image $\textbf{I}_F$ and is formulated as, 
\begin{equation}
\label{eqn:formula}
\phi(\textbf{x}) = \textbf{x} + u(\textbf{x}),
\end{equation}
where, $\textbf{x}$ represents voxels/pixels in the moving image $\textbf{I}_M$, $u(\textbf{x})$ denotes the displacement field, and $\phi(\circ)$ represents the deformation function.

To generate deformation fields that are locally smooth and discontinuous at the boundaries of different organs/regions, we propose to generate deformation fields for different sub-regions, and compose them to obtain the final deformation field. Sub-regions in the images to be registered must first be segmented either manually or automatically. With short-axis (SAX) cardiac cine-magnetic resonance (CMR) images, manual and automatic segmentation results for left ventricle blood pool (LVBP), left ventricle myocardium (LVM) and right ventricle (RV) are generally available in public data sets, large-scale imaging initiatives (e.g. UK Biobank) and from previous studies on automatic CMR segmentation~\cite{bai2018automated}. As the focus of this paper is on SAX-CMR image registration, we explicitly model discontinuities along cardiac boundaries by splitting the images into four sub-regions, namely, LVBP, LVM, RV, and background. These sub-regions are subsequently used to train our DDIR approach and register CMR images in manner that preserves discontinuities at their boundaries.

\begin{figure}[h]
\begin{center}
\includegraphics[width=\textwidth]{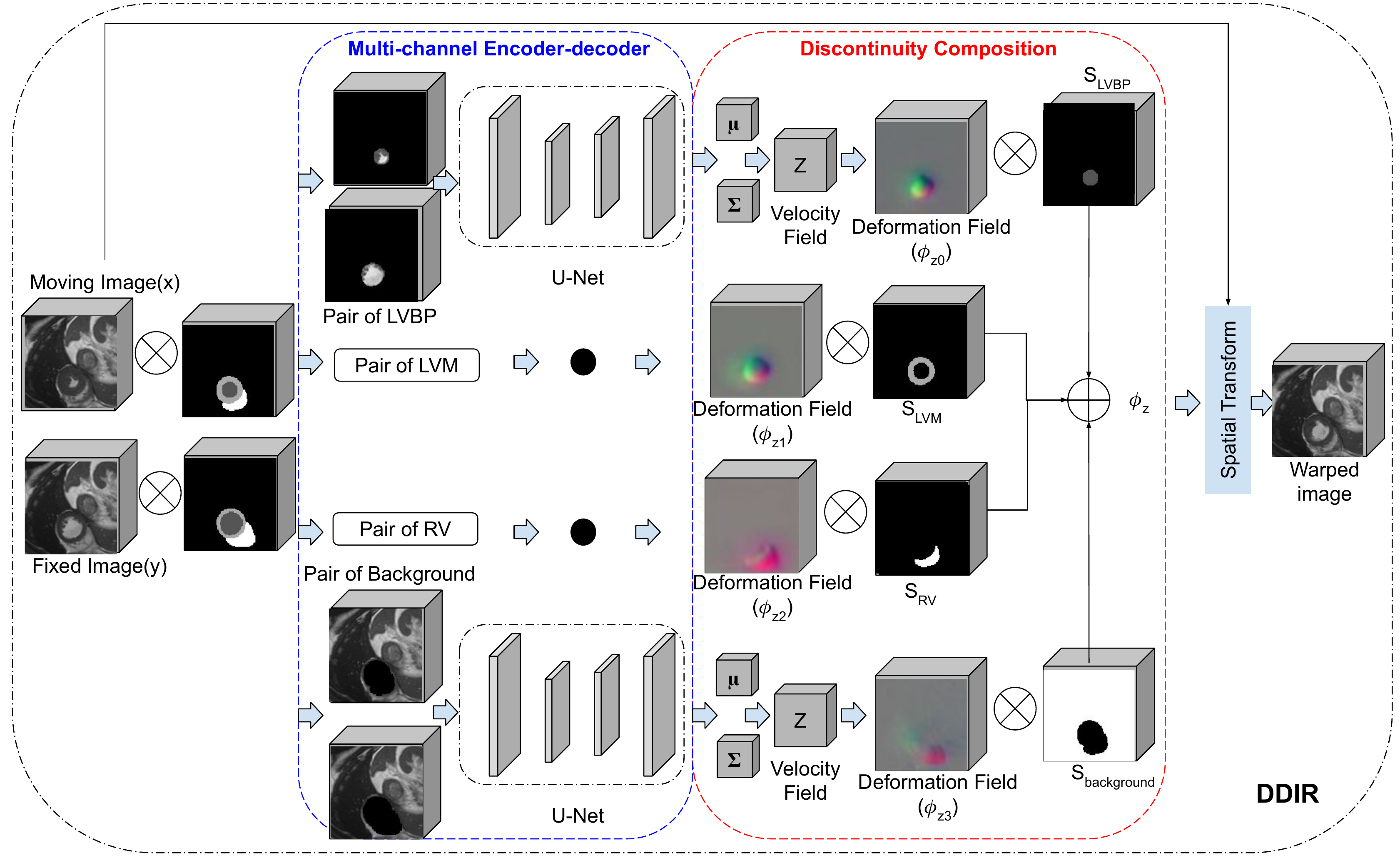}
\caption{Schema of DDIR. The registration network applies four different channels extracting features from pairs of LVBP, LVM, RV and background. Based on them, we obtain four sub-deformation fields for different regions. The final deformation field is obtained by composing these four deformation fields with corresponding segmentation. The cardiac MR images were reproduced by kind permission of UK Biobank ©.}
\label{fig:network}
\end{center}
\end{figure}
\textbf{Network Architecture.} Most previous DL-based registration methods apply an encoder-decoder network (generally U-Net~\cite{ronneberger2015u}) to extract feature maps from the concatenated input moving image and fixed image. However, as shown in Fig.~\ref{fig:network}, in DDIR the original moving image and fixed image (at $128\times128\times32$) are divided into four image pairs, i.e. LVBP, LVM, RV and background, using segmentation masks for the corresponding regions. In each of these pairs, voxels in corresponding regions are preserved while the rest are set at zero. Each pair is concatenated and fed as input to a distinct U-Net block, which extracts region-specific feature maps. These four U-Nets have the same architecture, including four down-sampling layers and three corresponding up-sampling layers. Using this multi-channel encoder-decoder structure, we obtain four sets of feature maps ($64\times64\times16$) corresponding to different sub-regions. We use the same U-Net architecture (with identical hyper-parameters) in all DL-based registration approaches investigated in this study. 

\textbf{Discontinuity Composition.} Using the region-specific feature maps learned by the U-Nets, we first predict four different smooth deformation fields (corresponding to each region) and then compose them to obtain the final deformation field, to preserve local smoothness and discontinuity at the interfaces. Similar to previous papers~\cite{dalca2019unsupervised,krebs2019learning}, we assume the transformation function (denoted as $\phi_z$) is parametrised by stationary velocity fields (SVF) ($z_i, i \in [0,3]$), which are sampled from a multivariate Gaussian distribution. With the predicted feature map, we compute the mean $\mu_i$ and variance $\Sigma_i$ of $z_i$ (using two different convolution layers). Based on them, four SVFs ($z_0,z_1,z_2,z_3$) corresponding to different regions (LVBP, LVM, RV and background) are sampled. With the corresponding integration layer and up-sampling layer, we obtain four diffeomorphic deformation fields $\phi_{z_0}$, $\phi_{z_1}$, $\phi_{z_2}$ and $\phi_{z_3}$. As before, we use region-specific segmentation masks to extract each region of interest from the obtained deformation fields (setting the remaining voxels to zero) and compose them to generate the final deformation field. Denoting the segmented regions of LVBP, LVM, RV and background as $S_{LVBP}$, $S_{LVM}$, $S_{RV}$  and $S_{background}$ respectively, the composition can be formulated as,
\begin{equation}
\label{eqn:formula}
\phi_z=\phi_{z_0}\times S_{LVBP}+\phi_{z_1}\times S_{LVM}+\phi_{z_2}\times S_{RV}+\phi_{z_3}\times S_{background}.
\end{equation}

\textbf{Loss Function.} The loss function includes two terms, a dissimilarity and a regularisation term. The former is the distance between the warped moving image and the fixed image, while, the latter constrains the estimated deformation fields to be locally smooth (i.e. within each region), to avoid unrealistic deformations. The dissimilarity loss in DDIR captures the dissimilarity on both images and segmentations. We use normalised cross-correlation (NCC) $L_{NCC}$ to evaluate the similarity between the warped moving image and the fixed image. Let warped moving image be $x$ and fixed image be $y$, $L_{NCC}$ is computed as,
\begin{equation}
\label{eqn:ncc}
{L}_{NCC} = 1- \frac{\sum_{i} (x_i-x_m)(y_i-y_m)}{\sqrt{\sum_{i}(x_i-x_m)^2} \sqrt{\sum_{i}(y_i-y_m)^2}}, 
\end{equation}
where, $x_i$, $y_i$ are the intensity of the i-th voxel in the warped moving image and fixed image, respectively, and $x_m$, $y_m$ are the mean intensities of the corresponding images. As the region-wise segmentation masks are available, we also compute the region-wise dice loss, denoted ${L}_{Dice}$ as in~\cite{milletari}.

To preserve discontinuity at the interfaces of the organs/regions while ensuring local smoothness, a global smoothness constraint is not enforced on the composed deformation field. The composition of different deformation fields preserves discontinuities at interfaces, therefore, we only need to guarantee the deformation field of each sub-region smooth. This is achieved by regularising each sub-deformation field. Following Voxelmorph-diff~\cite{dalca2019unsupervised}, we calculate the Kullback-Leibler (KL) divergence between the approximate posterior $q_{\psi}(z|\textbf{I}_F;\textbf{I}_M)$ and the prior $p(z)$ ($p(z) = \mathcal N(z;0,\Sigma_{z})$) of each velocity field $z$, formulated as,
\begin{equation}
\label{eqn:regularisation}
\begin{aligned}
&R = KL(q_{\psi}(z|\textbf{I}_F;\textbf{I}_M)||p(z|\textbf{I}_F;\textbf{I}_M)),\\
&L_R = \frac{1}{4}(R_{LVBP} + R_{LVM} + R_{RV} + R_{background}),
\end{aligned}
\end{equation}
where $R$ denotes the regularisation for each deformation field and $L_R$ is the combined regularisation term. The $q_{\psi}(z|\textbf{I}_F;\textbf{I}_M)$ is a multivariate normal,
\begin{equation}
\label{eqn:posterior}
q_{\psi}(z|\textbf{I}_F;\textbf{I}_M) = \mathcal N(z;\mu_{\textbf{z}|\textbf{I}_F,\textbf{I}_M},\Sigma_{\textbf{z}|\textbf{I}_F,\textbf{I}_M}),
\end{equation}
where $\mu_{\textbf{z}|\textbf{I}_F,\textbf{I}_M}$ and $\Sigma_{\textbf{z}|\textbf{I}_F,\textbf{I}_M}$ are the mean and variance of the distribution, learned by convolution layers. The complete loss function used to train the network is,
\begin{equation}
\label{eqn:total}
\begin{split}
{L}_{total} = \lambda_0 \times {L}_{NCC} + \lambda_1 \times L_{Dice}+\lambda_2 \times L_R, 
\end{split}
\end{equation}
where, $\lambda_0$, $\lambda_1$ and $\lambda_2$ are used to weight the importance of each loss term.

\section{Experiments and Results}

\textbf{Data and Implementation.} The registration performance of the proposed approach is evaluated on SAX-CMR images (spatial resolution at $\sim 1.8 \times 1.8 \times 10 mm^3$), available from UKBB. We chose images from 2,000 subjects at random, and used images at end-diastole (ED) and end-systole (ES) for intra-subject registration. Among these, 1,600 subjects' data was chosen at random for training DDIR, equating to 3,200 image pairs (ED-to-ES or ES-to-ED registration). Image pairs from the remaining 400 subjects were used for testing. All CMR images were resampled to $1.50 \times 1.50 \times 3.15 mm^3$ using bi-cubic interpolation, and cropped to a size of $128 \times 128 \times 32$ (with zero-padding for images with fewer than 32 slices). The region-wise segmentation masks for all CMR images were obtained automatically using the segmentation method proposed in~\cite{bai2018automated}. DDIR was implemented using Python and Keras on a Tesla M60 GPU machine. The Adam optimiser was used for training, with a learning rate of $1e-4$. The batch size was set to 2, and the hyper-parameters $\lambda_0$, $\lambda_1$ and $\lambda_2$ were set to $20, 200, 0.1$ (determined empirically), respectively. The source code will be publicly available on the Github $\footnote{\url{https://github.com/cistib/DDIR}}$.

\textbf{Quantitative Comparison and Analysis.} To demonstrate the superiority of our approach, we compare DDIR with both traditional registration and DL-based registration methods. For the former, we choose Symmetric Normalisation (SyN) registration (3 resolution level, with 100 iterations in each sampling level) in ANTS~\cite{avants2011reproducible}, and B-spline registration (max iteration step is 2000, sampling 6000 random points per iteration) in SimpleElastix~\cite{marstal2016simpleelastix}, for comparison. For the latter, DDIR is compared with Voxelmorph-diff~\cite{dalca2019unsupervised}. As DDIR uses segmentation masks during training and inference, it is a weakly-supervised registration method. For fair comparison, we build three weakly-supervised versions of Voxelmorph - VM-Dice, VM(img+seg) and VM-Dice(img+seg). VM-Dice uses a Dice loss $L_{Dice}$ term and binary cardiac segmentation masks for the fixed and moving images during training, but does not require the latter for inference. In VM(img+seg), we concatenate the fixed and moving images with their corresponding multi-class masks (i.e. distinct labels for each region) and use these to train the network. While, VM-Dice(img+seg) is a combination of the previous two methods. We did not compare with the DL-based discontinuity-preserving method proposed in~\cite{ng2020unsupervised}, as there is no corresponding source code publicly available. This strategy to register different sub-regions and compose corresponding deformation fields is also applicable to the aforementioned networks. Hence, we also apply this strategy during inference, for trained Voxelmorph-diff and VM-Dice models (as they only require sub-images as input on the inference), for comparison with DDIR. These are denoted Voxelmorph-diff(compose) and VM-Dice(compose). These two approaches are different to DDIR as the composition of sub-deformation fields is not learned end-to-end during training (as in DDIR). 

To demonstrate the advantage of incorporating discontinuity in the DL-based registration network, we also build a baseline for DDIR, DDIR(baseline), where the predicted feature maps from the four different channels are concatenated and used to compute a single diffeomorphic deformation field (instead of four sub-deformation fields, as in DDIR).

\begin{figure}[b!]
\begin{center}
\includegraphics[width=\textwidth]{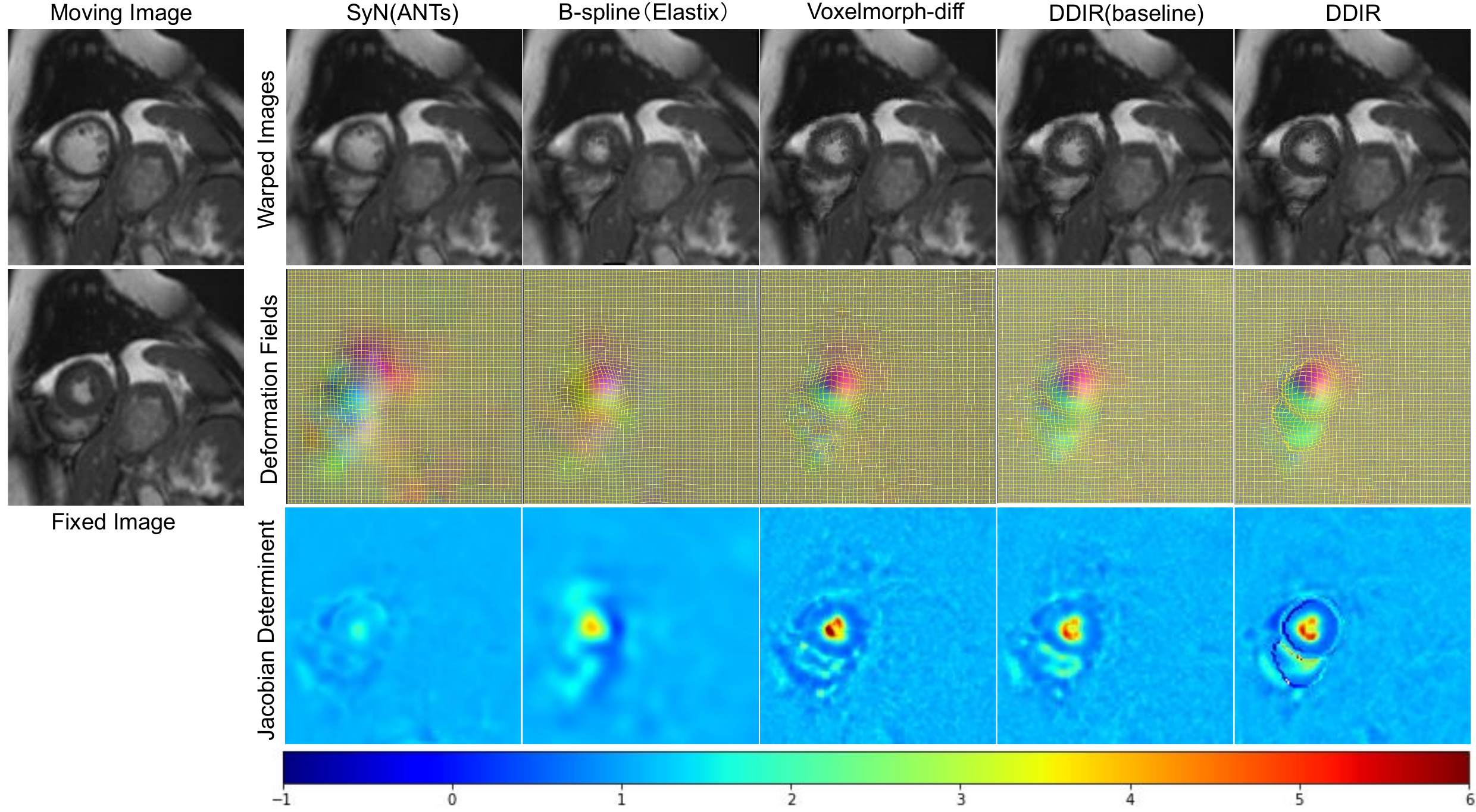}
\caption{Visual comparison of deformation fields estimated using DDIR and state-of-the-art methods. Left column: Moving and fixed images; Right column: corresponding warped moving image (first row), deformation fields (second row) and Jacobian Determinant (last row). Colours in the Jacobian determinant images, from blue to red represent the intensity from low to high. The cardiac MR images were reproduced with permission of UK Biobank ©.}
\label{fig:qualitative_results}
\end{center}
\end{figure}

\textbf{Qualitative Results.} Registration results obtained using DDIR and the other methods investigated are assessed visually in Fig.~\ref{fig:qualitative_results}. Here, the moving and fixed images are shown in the first column. The corresponding warped moving images, deformation fields, and Jacobian determinants (rows 1-3) obtained following registration using SyN, B-spline, Voxelmorph-diff, DDIR(baseline) and DDIR, are shown in columns 2-6. The warped moving images obtained by both traditional registration methods distinctly different to fixed image, although the B-spline result appears visually more similar than obtained by SyN. All warped moving images obtained using DL-based methods look more similar to the fixed image, than the former. The deformation fields and their corresponding Jacobian determinants estimated using each approach indicate that distinct boundaries for the left and right ventricle are retained using DDIR, not afforded by the rest.

\begin{table*}[h] 
 \caption{\label{tab:comparison} Quantitative comparison between DDIR and state-of-the-art methods using the DS of LVBP, LVM, RV and average Dice (denoted as Avg. DS) and HD. Statistically significant improvements in registration accuracy (DS and HD) are highlighted in bold. Besides, LVEDV and LVMM indices with no significant difference from the reference are also highlighted in bold.} 
 \centering
 \resizebox{12cm}{!}{
 \begin{tabular}{lccccccc} 
 \hline
  Methods  & LVBP DS (\%) & LVM DS (\%) & RV DS (\%) & Avg. DS (\%) & HD (mm)& LVEDV & LVMM \\ 
 \hline
  before Reg & $57.68 \pm 6.21$ & $30.88 \pm 8.68$ & $55.13 \pm 7.51$ & $47.90 \pm 6.33$ & $12.91 \pm 2.48$ & $143.76 \pm 32.13$ & $83.67 \pm 21.06$\\ 
  B-spline & $74.44 \pm 11.50$ & $68.06 \pm 7.20$ & $61.76 \pm 12.05$ & $68.09 \pm 8.76$ & $13.72 \pm 3.57$ & $131.14 \pm 40.64$ & $81.11 \pm 22.60$ \\ 
  SyN & $70.92 \pm 9.36$ & $57.88 \pm 10.59$ & $60.30 \pm 8.35$ & $63.03 \pm 8.29$ & $12.98 \pm 2.68$  & $120.09 \pm 41.83$ & $\textbf{83.12$\pm$ 21.20}$ \\ 
  Voxelmorph-diff & $81.73 \pm 8.71$ & $72.04 \pm 4.65$ & $65.73 \pm 9.62$ & $73.16 \pm 6.26$ & $12.96 \pm 3.14$ & $137.16 \pm 32.59$ & $78.65 \pm 21.68$ \\ 
  VM-Dice & $82.28 \pm 8.75$ & $72.53 \pm 4.59$ & $66.30 \pm 9.67$ & $73.70 \pm 6.28$ & $13.00 \pm 3.24$& $139.58 \pm 32.79$ & $78.98 \pm 21.57$\\ 
  VM(img+seg) & $82.54 \pm 8.50$ & $72.66 \pm 4.80$ & $66.69 \pm 9.64$ & $73.96 \pm 6.28$ & $12.68 \pm 3.21$& $138.29 \pm 33.00$ & $80.83 \pm 21.62$ \\
  VM-Dice(img+seg) & $81.97 \pm 8.53$ & $71.23 \pm 4.79$ & $70.20 \pm 12.05$ & $74.47 \pm 6.79$ & $11.28 \pm 4.35$& $\textbf{144.33 $\pm$ 32.93}$ & $80.17 \pm 22.02$\\
 \hline
 Voxelmorph-diff(compose)& $78.82 \pm 6.38$ & $67.41 \pm 8.80$ & $75.10 \pm 6.97$ & $73.78 \pm 6.10$ & $11.74 \pm 3.08$ & $119.30 \pm 38.71$ & $91.39 \pm 23.07$\\ 
 VM-Dice(compose)& $79.59 \pm 5.91$ & $68.81 \pm 7.81$ & $\textbf{77.93 $\pm$ 6.63}$ & $75.44 \pm 5.36$ & $11.14 \pm 3.12$ & $120.90 \pm 38.14$ & $94.89 \pm 25.96$\\ 
 \hline
 DDIR(baseline) & $84.25 \pm 8.63$ & $75.02\pm 4.50$ & $71.42\pm 10.32$ & $76.90 \pm 6.58$ & $11.85 \pm 3.38$ & $\textbf{141.73 $\pm$ 32.29}$ & $79.01 \pm 21.40$\\
 DDIR & $84.63 \pm 8.07$  & $75.27\pm 5.03 $ & $74.07 \pm 8.73 $ & $\textbf{77.99 $\pm$ 5.47}$ & $\textbf{10.65 $\pm$ 3.51}$ & $\textbf{141.84 $\pm$ 32.59}$ & $\textbf{81.92 $\pm$ 21.86}$ \\
 \hline
 \end{tabular}
  }
\end{table*}

\textbf{Quantitative Results.} To quantitatively evaluate the performance of our approach, we compare DDIR with previous methods using Dice score (DS) and the Hausdorff Distance (HD). DS is computed for LVBP, LVM and RV. These values and the average DS and HD across all regions are reported in Table~\ref{tab:comparison}. Besides, to demonstrate the clinical value of DDIR, we also compute two clinical indices, LV end-diastolic volume (LVEDV) and LV myocardial mass (LVMM). The former is computed using ED segmentations, while the latter, is computed using ED and ES segmentations, pre- and post-registration. Pre-registration, LVEDV and LVMM are computed based on the moving and fixed segmentations (used as reference values). Post-registration, we compute them based on the warped moving segmentation. Therefore, as we perform both ED-to-ES and ES-to-ED registration for each subject, the LVMM values reported in Table~\ref{tab:comparison} represent the average computed at both ED and ES, across all subjects. Thus the closer LVEDV and LVMM (post-registration) are to the reference values, the better the registration performance.

DL-based approaches outperform traditional registration methods in terms of both DS and HD. The weakly-supervised variants of Voxelmorph-diff provide improvements over Voxelmorph-diff, consistent with previous research\cite{dalca2019unsupervised}. Using segmentation masks as additional input channels to the network (VM(img+seg)) yields better results than using them just to compute the loss and drive gradient updates (VM-Dice) (73.96\% vs 73.70\%). However, conversely the former requires segmentation masks during inference, while the latter do not. The combination of these two strategies (VM-Dice(img+seg)) further improves registration performance ($\sim0.5\%$ in terms of average DS). Composing sub-deformation fields also improves registration accuracy of the trained networks, with Voxelmorph-diff (compose) achieving $0.6\%$ higher average DS than Voxelmorph-diff (73.78\% vs 73.16\%), and VM-Dice (compose) achieving $\sim1.7\%$ higher average DS than VM-Dice (75.44\% vs 73.70\%).

We found that the DDIR(baseline) achieves $\sim1\%$ higher average DS than VM-Dice(img+seg) (76.90\% vs 75.93\%), which highlights the advantage of using a multi-channel encoder-decoder network. Compared with DDIR, we found that incorporating discontinuity further improves the average DS (77.99\% vs 76.90\%). Correspondingly, DDIR also obtains the best performance in terms of the DS for LVBP, LVM and HD, while its RV DS is lower than VM-Dice(compose). We evaluated the statistical significance of these results using paired t-tests and found that DDIR significantly outperforms Voxelmorph-diff, VM-Dice, VM(img+seg) and VM-Dice(img+seg) on all DS and HD metrics (P-value$<$0.05). DDIR also significantly outperforms DDIR(baseline) in terms of average DS, RV DS and HD.
Each sub-deformation field generated by DDIR are smooth (without foldings). After composing, the discontinuity only exists at the interface of different sub-regions, which demonstrates that DDIR can generate locally-smooth but globally-discontinuous deformation fields.

The clinical indices, LVEDV and LVMM, show no significant differences (P-value$>$0.05) post-registration using DDIR to the reference values, not afforded by other approaches. This demonstrates the superiority and clinical value of our method. To analyse the discontinuity on the deformation fields, we visualise the deformation fields generated using DDIR and DDIR (baseline) (presented in the supplementary material), where the discontinuity is observed for the former along the LV and RV boundaries.

\section{Conclusion}
We proposed a novel weakly-supervised discontinuity-preserving registration network, DDIR, which significantly outperformed the state-of-the-art, in intra-patient CMR registration. DDIR preserves LV clinical indices post-registration, not afforded by the other approaches. This makes it compelling as a tool for use in clinical applications as it ensures that common diagnostic biomarkers for the LV are preserved post-registration.

\section*{Acknowledgements}
This research was conducted using the UKBB resource under access application 11350. 

\bibliographystyle{splncs04} 
\bibliography{ddir}
\end{document}